\documentclass{article}
\usepackage{fullpage}
\usepackage{graphicx}
\usepackage{longtable}

\usepackage {listings}
\usepackage{array,booktabs}
\newcolumntype{L}{@{}>{\kern\tabcolsep}l<{\kern\tabcolsep}}
\usepackage{colortbl}
\usepackage{xcolor}

\begin{document}
This is a pre-print of the following Chapter: Arvind W. Kiwelekar,  Pramod Patil Laxman D. Netak and Sanjay U Waikar, {\em Blockchain-Based Security Services for Fog Computing} accepted and final version  is published in  Chang W., Wu J. (eds) Fog/Edge Computing For Security, Privacy, and Applications. Advances in Information Security, vol 83. Springer.

{\bf Cite this chapter as:}
Kiwelekar A.W., Patil P., Netak L.D., Waikar S.U. (2021) Blockchain-Based Security Services for Fog Computing. In: Chang W., Wu J. (eds) Fog/Edge Computing For Security, Privacy, and Applications. Advances in Information Security, vol 83. Springer, Cham. 
\newpage
  
\title{Blockchain-based Security Services for Fog  Computing}

\author{Arvind W. Kiwelekar, Pramod Patil, Laxman D. Netak, Sanjay U Waikar \\
 Department of Computer  Engineering\\
Dr. Babasaheb Ambedkar Technological University\\ 
Lonere, Raigad-402103, India \\
\{awk,ldnetak\}@dbatu.ac.in, patilpramod042@gmail.com, sanjaywaikar@yahoo.com}

\maketitle

 \begin{abstract}
Fog computing is a paradigm for distributed computing that enables sharing of resources such
as computing, storage and network services. Unlike cloud computing, fog computing
platforms primarily support {\em non-functional properties} such as location awareness, mobility
and reduced latency. This emerging paradigm has many potential applications in domains such as smart grids, smart cities, and transport management.

Most of these domains collect and monitor personal information through edge devices to offer personalized services.   A {\em centralized} server either at the level of cloud or fog, has been found ineffective to provide a high degree of security and privacy-preserving services. 

Blockchain technology supports the development of {\em decentralized} applications designed around the principles of immutability, cryptography, consistency preserving consensus protocols and smart contracts. Hence blockchain technology has emerged as a preferred technology in recent times to build trustworthy distributed applications.

The  chapter  describes the potential of blockchain  technology to realize  security services such as
authentication, secured communication, availability, privacy and trust management to support the development of dependable fog services.
 \end{abstract}

\section{Introduction}
Emerging technologies are impacting our lives in two different ways. First, these technologies are improving our {\em standard of living}.  For example, Artificial and Machine Learning are the technologies behind personalized health care, intelligent transport services,  free and open education to all. Second, they are also improving the {\em quality of service} we expect from service providers. Technologies such as the internet and mobile communication are providing the quality of services which was unimaginable a few years back. For example,   these technologies enable 24 X 7 banking services, global-market for selling local products, and opportunities to monetize excess personal resources through aggregated services like Airbnb.

In this chapter, we review the impact of two such emerging technologies called Blockchain Technology and Fog Computing. Both technologies improve the standard of living and the quality of services offered to us through the internet.

Diverse domains such as  Crowd Surveillance and Public Safety, Geospatial Data Analysis, Intelligent Transport Service, Smart Grid and Smart Healthcare have started adopting Fog Computing in recent times.  Adoption of Fog Computing mainly aims to reduce the response time required for accessing critical services like energy, healthcare and transportation.

Deep penetration of information and communication technologies in our social life is also raising concerns about the security and privacy of the personal data collected through them.  In recent times, use of  Blockchain technology has increased for protecting personal data so that trustworthy system can be built.

The chapter contributes by presenting an evaluation of Blockchain Technology in the context of Fog Computing. We first identify the security requirements for various application domains of Fog Computing. Then we present a detailed analysis of the strengths and weaknesses of Blockchain Technology to meet these security requirements.

Rest of the chapter is organized as below: (i) A brief overview of fog computing (ii)
Fog computing use cases and their Security
requirements (iii)  Generic Security requirements for Fog Applications (iv) A Blockchain Primer. (v) Blockchain-based Security Solutions (vi) Conclusion.

\section{Fog Computing: Introduction}

The Fog Computing (FC) \cite{satyanarayanan2009case} is emerging as a complementary computing paradigm for Cloud Computing (CC) to meet the computing, storage, and network requirements of {\em resource-constrained} computing nodes.  
Smartphones, tablets,   the Internet of Things (IoT), wireless sensors and actuators are some of the examples of {\em resource-constrained} computing devices. Such kinds of devices have limited computing power, small memory, and access to the network through wireless media. Despite their limited capacity, such types of devices are transforming the nature of computing from an enterprise phenomenon into a pervasive phenomenon.

In this section,  we describe limitations of CC to meet the requirements of resource-constrained devices followed by a description of distinct characteristics of the Fog in comparison with the Cloud.

\subsection{Limitations of Cloud Computing}

The Cloud Computing (CC)  is now an established alternative to meet the computing, storage and networking requirements of enterprises in the presence of the reliable Internet.  The cloud provides computing resources and services to remote machines on a {\em pay-per-use} billing model.   Additionally,  the CC  environment offers flexible deployment models such as Platform as a Service(PaaS,e.g., Google's Cloud Services), Infrastructure as a Service(IaaS, e.g., Amazon's Elastic Computing Cloud), and Software as a Service (SaaS, e.g., Salesforce's Cloud Services).  This flexibility makes CC  a cost-effective solution to host resources and services for enterprise computing needs \cite{10.1145/2756545}.  

The CC paradigm has been found useful especially for enterprise resource planning \cite{saini2011cloud}, customer relations management, e-business owing to its characteristics such as high scalability, ease-of system administration, and support for rich programming models. 

However, the CC environment falls short to meet various requirements of  {\em resource-constrained devices} which include IoT,wearable devices, wireless sensors and actuators. Some of these requirements identified in the Reference \cite{puliafito2019fog} are described below.  

(i) {\em Latency:}  Video streaming, gaming, smart factories, and connected vehicles are some of the application scenarios which use devices like IoT and wireless sensors.  The latency requirements of such applications fall in the range of microseconds to tens of milliseconds.  The average latency experienced by resource-constrained devices when they are connected to the cloud falls in the range of hundreds of millisecond.   This high latency is undesirable in such application scenarios.

(ii) {\em Bandwidth:}  The resource-constrained devices typically access the network through a wireless medium. At the same time, applications enabled by these devices such as smart factories produce data at the rate of thousands of gigabyte per second. The cloud computing environments fall short to meet such high bandwidth requirements. 

(iii) {\em Privacy and Security:} In some of the application contexts such as health monitoring and control, devices need to transmit private and personal information for remote processing. The resource-constrained devices lack the computing power to execute complex encryption algorithms needed to secure data when it is transmitted over the public Internet as in the case of cloud computing.  Hence, securing such information becomes a challenge when resource-constrained devices are connected to the cloud.

(iv) {\em Context Awareness:}   In application scenarios such as connected vehicles, Intelligent Transport Systems(ITS) need to transfer context information. For example, information about traffic conditions, weather information, location and information. When resource-constrained devices are 
 connected to a distant cloud data centre, transmitting such local information has little temporal and spatial relevance.

\subsection{Distinct Features of Fog Computing}

From the functional point of view, Cloud Computing and Fog Computing are similar phenomena which provide computing, networking and storage resources to remote machines. Both environments include resource-rich devices such as high-end servers accessed through either public or private networks. Although the business model for the FC is currently evolving, similar to CC, the business model of the FC  in future may be centred around pay-per-use billing mechanisms and hosting of resources by a third party.

\begin{table}[t]
    \centering
    \begin{tabular}{|p{1.3in}|p{1in}|p{1in}|p{1in}|}
    \hline
    
      Feature   &  Edge Computing & Fog Computing & Cloud Computing \\ \hline
      Latency & Low & Medium & High  \\ \hline
    Bandwidth & Low & Medium & High  \\ \hline
Compute capacity & Low & Medium & High  \\ \hline
Reliable compute & Low & Medium & High  \\ \hline
Reliable connectivity  & Low & Medium & High  \\ \hline
Data longevity & Low & Medium & High  \\ \hline
    \end{tabular}
    \caption{A Comparison of Edge, Fog and Cloud Computing \cite{varshney2017demystifying}}
    \label{comptab1}
\end{table}

In terms of software engineering terminology, Fog computing and Cloud computing differ regarding non-functional requirements. It includes Performance requirements, Reliability requirements, Deployment models and Security requirement. These requirements are also known as operational requirements. Table \ref{comptab1} shows a comparison of Edge, Fog and Cloud Requirement \cite{varshney2017demystifying}.

Hence, to handle these non-functional requirements emanating from the requests of edge devices, a new computing paradigm has emerged in recent times called Fog Computing. The Fog Computing which has introduced a new application management layer in the middle between cloud and edge devices referred to as a Fog layer. The Fog layer extends the cloud management services and brings them nearer to the network.

 Fog and Cloud mainly differ in terms of latency.  The latency to transfer data from a Fog to edge devices is lower than when data transfer occurs from an edge device to a Cloud. This lower latency is because of edge devices  are a one-hop topological distance from fog servers. Also, the network bandwidth between edge devices and the Fog is much higher through a wireless link than between edge devices and the Cloud.
 
Additionally, the Fog stores the data transferred from edge devices for a shorter period; the Fog periodically pushes the data to the Cloud for archival purposes.

Mobility is another distinct non-functional parameter in which Cloud and Fog Computing differ. The servers and computing nodes hosting cloud management services are centralized one. When they are geographically distributed, often the computing nodes reside in an office premise and not mobile.  Unlike this configuration, a Fog may host computing nodes and services in mobile vehicles. Also, the number of requests that a Fog may have to handle from mobile clients are enormous.

Additionally, it is also essential to know the differences between Fog Computing and Edge Computing.  Although, the differences between fog and edge computing are blurred one, we discuss here some of them.  An Edge Computing node supports the computing requirement of edge devices which include wireless sensors and actuators. Edge computing nodes are directly interfaced with edge devices. An edge computing node communicates with edge devices through conventional communication mode such as pooling and interrupts in contrast to client-server communication used in Fog Computing. The edge computing node supports hardware-enabled security, unlike application-level security provided in Fog Computing. Further, the Edge Computing nodes typically use flash storage devices, unlike spinning storage disks used in Fog Computing.

\section{ Fog computing use cases and their Security requirements}
  Many application domains such as listed in Table \ref{secreq}  have started adopting  Fogs over Clouds to meet their computation,  storage and networking requirements. For these application domains,  Fog platforms meet their requirements of low-latency, high bandwidth and context-awareness. At the same time, these application domains have stringent security requirements. A brief description of the security requirements specific to these domains follows.

\begin{table}[t]
    \centering
    \begin{tabular}{|p{0.25in}|p{1.5in}|p{2.75in}|}
    \hline
       Sr.  &  Application Domains & Security Requirements  \\ \hline 
        1  & Urban Surveillance and Public Safety   & Privacy and Autonomy \cite{elmaghraby2014cyber}, Panoptic Systems  \cite{van2011policing,xu2019blendmas}\\ \hline
        2  &  Smart Power Grid &  Denial of Service Attacks \cite{bou2013communication}, Integrity Attacks \cite{metke2010security}, Malware attacks, Power thefts, Billing Manipulations \cite{aloul2012smart} \\ \hline
        3  & Geospatial Data Analysis (UAV)  &  Secured Communication \cite{javaid2012cyber,he2016communication},  Man in the Middle attack,  Privacy \cite{rodday2016exploring} \\  \hline
        4  & Intelligent Transportation Systems  (ITS) and  Connected Vehicles  &  Authentication, Availability, Non-Repudiation, Integrity \cite{siegel2017survey},
         Denial of Service, Sybil, Black-hole attack \cite{sakiz2017survey}
        \\ \hline
        5  & Smart Healthcare &  Data Confidentiality, Data Authentication,  Data Integrity, Availability  for wireless body network \cite{saleem2009security}    \\ \hline 
       6  & Industry 4.0 &  Enterprise Cyber-Espionage,denial of service attacks, and phishing attacks \cite{pereira2017network}   \\ \hline 
         
    \end{tabular}
    \caption{Security Requirements for Fog Computing Use-cases}
    \label{secreq}
\end{table}

\begin{enumerate}
    \item {\bf Urban    Surveillance    and Public Safety} 
Low-cost surveillance technologies such as CCTVs and sensors enable to collect and monitor data about people living in urban areas. For example,  law enforcement agencies can track the movement of suspicious people in designated sensitive areas to prevent any public damage and crimes.  
The collected data is location-specific and relevant to take timely decisions.  Hence the fog computing paradigm is an appropriate alternative for storage and analysis purposes.

Though the data is collected to provide public-safety, it is susceptible for misuse either by the fog service providers or edge operators who transmit the surveillance data.   One of the frequently cited threats includes a Fog node operator may share the collected information about the movement of a person to a third party without informing the concerned person. Another example of threat includes denial of service attacks through flooding the network by malicious edge operators.   At the same time, such systems are giving rise to a panoptic system which continuously monitors citizens.

    \item {\bf Smart Power Grid} In the energy sector, the increased thrust upon the adoption of renewable energy sources (e.g., solar, wind) has changed the relationship between energy generators and consumers. The conventional energy systems are mostly fuel or coal-based, centralized, and information flows from the generator to the consumer.  The modern energy sector is increasingly using renewable energy sources, a large number of energy distributors are dispersed along a wide geographical area, and the information flows in both the direction. The network of energy generators, distributors and consumers called smart grid \cite{liserre2010future} is formed through the use of information and communication technologies, sensors, and actuators to effectively operate the energy grid. 
    
    To effectively operationalize smart grids, Fog computing has emerged as a preferred distributed paradigm in comparison with cloud computing in recent times \cite{okay2016fog}. The guaranteed response time, a large number of decentralized grid operators and stringent privacy requirements from the consumer point of view are some of the factors behind the preference of fog computing over cloud computing.

The security requirements in Smart grid arise from the domain-specific concerns such as assuring the integrity of the data communicated between grid operators and consumers \cite{bou2013communication}.  This data includes valuable information such as billing information, and, energy usage patterns of consumers. 
Further, a malicious smart meter can overload the network to disrupt and deny services to authorized customers from accessing the services provided by a Fog service provider.

    \item{\bf Geospatial Data Analysis} : Low-cost technologies such as Unmanned Aerial Vehicles (UAV), Radio Frequency Identifiers (RFID) and GPS enabled devices are producing a large amount of geospatial data \cite{lee2015geospatial}. Geographic Information Systems (GIS) manage and analyze such geospatial data to support urban planning, agriculture and environment monitoring. 
    
    The requirements for reduced storage space, reduced transmission power, reduced latency and increased throughput are driving software engineers to adopt the Fog computing paradigm to build  GIS applications \cite{barik2016foggis}.
    
    The geospatial data need to be protected from different types of security attacks to ensure regional security and privacy of persons who share the data.  
The commonly employed techniques are trust management in GIS service provider, data integrity checks, and authentication of GIS users \cite{bertino2008security}.
    
    \item {\bf Intelligent   Transportation Systems (ITS) and  Connected Vehicles}
    The   Intelligent Transport System (ITS) refers to the use of Information and Communication Technologies  (ICT) for improving the efficiency and effectiveness of transport services.  Some of the technologies that form the backbone of ITS are Wireless Sensors and actuators, Cloud Computing,  and GPS controlled vehicles \cite{perallos2015intelligent}.   The Connected Vehicle (CV) is another related concept that is enabling the evolution of the next generation of ITS and Internet of Vehicles(IoV). The connected vehicle refers to using wireless technologies for communicating with other vehicles and the infrastructure offering transport services\cite{lu2014connected}.
    
     The ITS and Connected Vehicle have started utilizing the advantages of Fog Computing such as scalability, low latency,  and context awareness to improve the Quality of Services. The use of Fog Computing for ITS reduces the average trip time,  CO2  emissions and fuel consumption
     \cite{brennand2016fox}.
     
     Jin Cui et al. identify and catalogue various kinds of security attacks for which autonomous vehicles and  ITS  need to protect. These include authentication, availability, data integrity, confidentiality and privacy\cite{cui2019review}.
    
    \item {\bf Smart Healthcare}
    To make healthcare more personalized and precise, medical systems have started adopting technologies such as wearable health monitoring devices, IoT, big data analysis, and Artificial Intelligence. Such health care systems, referred to as smart healthcare systems, have to address computational and security challenges.

In the context of smart healthcare, the Fog-based platforms tackle the computational challenges  by bringing resources closer to the patients, reducing response time and by providing energy-efficient data processing\cite{ahmad2016health}.

Preserving the privacy of the patient's data and making health care services  available round the clock are some of the security challenges that need to be addressed effectively \cite{saleem2009security}.

    \item {\bf Industry 4.0}
    The combination of ICT, IoT and intelligent systems have revolutionized manufacturing and production systems in recent times. This industrial revolution is named as Industry 4.0 \cite{lee2015cyber}.  Industry 4.0 has brought a  transformation into the nature of manufacturing units from the automated one to an autonomous one.  

The Fog Computing is a technology that is leading this 4th industrial revolution because of its inherent strengths such as low latency rate \cite{o2018fog}, low power consumption and proximity to wireless sensors and actuators which monitor and control various production processes.

Some of the common security attacks observed in smart manufacturing systems are: (i) the leakage of critical production information, and (ii) withholding access to a manufacturing unit. These security threats intend either to disrupt the production process or the production schedule \cite{pereira2017network}. 
\end{enumerate}

\section{Generic Security requirements for Fog Applications} \label{genreq}

The previous section briefly surveys  security requirements for various use cases of Fog Computing. Some of the security requirements are common across more than  one application domains. For example, protecting end users from the {\em denial of service attacks} is a requirement of   ITS,  Industry 4.0, and other domains.  This section identifies and explains such generic requirements common across various Fog applications. 
\begin{enumerate}
 \item {\bf Authentication} Authentication is the primary service in distributed and networked environment. The purpose of authentication is to verify and validate the identity of end users. An end user may be a person  or a device or an application who would like to access a service.   The task of authenticating is a primitive operation because it ensures that only legitimate users can enter the network.
 
 Some of the mechanisms that are commonly used for authenticating  users in cloud computing are: passwords, hard/soft tokens, device identification, bio-metric identification or a combination of these techniques \cite{ziyad2014critical}.
 
 While devising  effective authentication services for Fog Computing constraints such as  resource limitations of edge devices, high mobility of fog nodes and edge devices, network heterogeneity and availability of wired  wireless communication need to be considered \cite{ziyad2014critical}.

    \item {\bf Secured Communication} Assuming a fool-proof  underlying secure communication channel leads to many security attacks such as eavesdropping, spoofing,and information leakage at application level. Hence,  Cloud as well as Fog applications need  to protect the integrity of application data by providing a secured communication channel on top of underlying un-secured medium.
    
    Two types of communications are observed in Fog networks. First, a communication between edge devices and fog nodes.  This communication can be secured through symmetric key cryptography.  Maintaining an public key infrastructure and reducing message overhead are some of the challenges that need to be addressed considering resource constraints of Fog networks.
    
    Second,  providing end-to-end security in the presence of multiple hops in a fog network and mobility of fog nodes are some of the challenges that need to considered while securing communication among fog nodes.

    \item {\bf Availability} One of the critical requirements that is common across the domains is that the services offered as Fog services need to be made available round the clock.  Malicious users adopt techniques  such as flooding the network with illegal packets or re-routing network traffic to a  wrong destinations for denying requested services to  legitimate  users. Promptly detecting  and protecting against such threats can save lives in Health and IIS domains.
    \item{\bf Privacy}
    Most of the Fog applications track personal information to provide personalized services. Few examples follow. First, systems like ITS and urban surveillance monitor mobility patterns of citizens which have personal value. Second,  in case of smart grids, energy usage patterns are tracked and monitored by grid operators. Third, smart healthcare systems store personal and medical history of patients. Privacy is at stake when service providers use such critical personal information  for monetary gains or for competitive advantage without the consent of service users. Designing fog layer which protects unintended usage of such personal information  is a challenge. 
    
    \item {\bf Trust Management} Trust in network-centric systems is a bidirectional phenomenon. Service providers need to earn the trust of service users by providing timely and  secure responses. Also, service users need to demonstrate to service providers that they are the legitimate and non-malicious users. Such bidirectional trust is built through a series of interactions among service providers and users.
    
    Quantifying reputations of service providers,  opinions of  service users and service level agreements are some of the techniques used in case of cloud-based service providers.  
    
    Dynamic nature of fog nodes i.e. a fog node leaves and joins network dynamically and mobility of edge devices are some of
   the factors  which need to be considered while implementing a trust management system at Fog layer.
    
\end{enumerate}

The emerging blockchain technology has potential to address these security concerns in the context of Fog Computing. Before discussing blockchain-based solutions, we are briefly reviewing the essential elements of blockchain technology follows.

\section{A Blockchain Primer}

The necessity of blockchain technology can be understood  by evaluating   potential and pitfalls of the Internet as a platform for business.

The Internet has introduced an information-centric model of business, and it has revolutionized the way people transact online. For example, the emergence of e-commerce sites (e.g., Amazon) has been attributed to the growth and widespread presence Internet.

The Internet has bridged the information gap that exists between a service provider and service consumer by creating a third-party for information exchange called intermediaries or agents or service providers. These agents which are e-commerce sites, hold the information about who sells what, i.e. seller's information and who wants what, i.e. buyers profile and their needs — thus bringing together consumers of services or goods with that of producers.

Some of the advantages of doing  business online include  the process of business transactions is simplified, and   the time required for businesses is reduced. 

Despite the various benefits of the Internet, it has always remained an unreliable platform to share valuable personal information because of its mediator-centric model for information exchange.  A server or mediator may be a payment gateway or an e-commerce site. The information shared with such sites is always susceptible to breach of security and privacy attacks.

The emerging blockchain technology removes these pitfalls by laying a trust layer on top of the existing Internet technology. It replaces the mediator-centric model of information exchange with the peer-to-peer model or decentralized model.   It transforms the Internet into a trustworthy platform for doing business when transacting parties do not trust each other. It eliminates the role of mediator responsible for authenticating the identities of transacting parties. 
Initially emerged as a platform to exchange digital currency over the Internet, now the blockchain technology is gradually emerging as a general-purpose platform for sharing and protecting information.  

The four fundamental concepts common across the blockchain implementation are \cite{dinh2018}:
(i) Distributed Ledger,
(ii) Cryptography,
(iii) Consensus Protocols, and
(iv) Smart Contracts

\subsection{Distributed Ledger} 

In a conventional sense, ledgers are the registers or
logbooks employed for account-keeping or book-keeping 
operations. Similarly, in the context of a blockchain-based
information system, ledgers are the databases storing
up-to-date information about business transactions. These
are distributed among all the nodes participating in the
network.  So multiple copies of a ledger exist in a
business network.  When a node in a network updates its
local copy, all other nodes synchronize their copy with the
updated one.  Hence, each copy is consistent with each
other.

These ledgers are used to store information about valuable assets. In the Bitcoin implementation, the first blockchain-based system, ledgers are used to store digital currencies. It may be used to store information about other valuable assets such as land records, diamonds, student's academic credentials and others.

\begin{figure}[h]
\centering
    \includegraphics[width=4.5in]{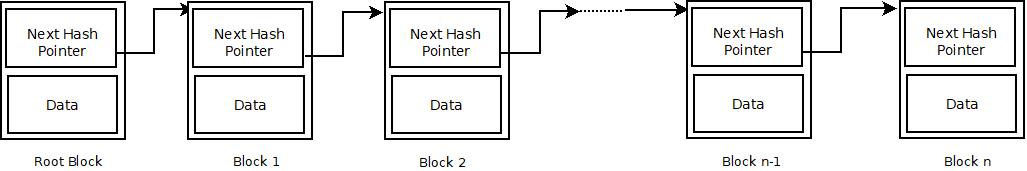}
    \caption{Blockchain}
    \label{bc}
\end{figure}

In a blockchain-based information system, records in a distributed ledgers are arranged in a chain-format, as shown in Figure \ref{bc} for storage purpose. Here, multiple transactions related to an asset are grouped in a block. The $(n+1)^{th}$ block in the chain links to the $n^{th}$ block and the $n^{th}$ block links to the $(n-1)^{th}$ block and so on.  Due to this peculiar storage arrangement, the distributed ledgers are also known as Blockchain.  The blockchain data structure permits only append of new records. Updating and deletion of records are not permissible.

The most critical design feature of blockchain-based information system is the use of hash pointers instead of physical memory based pointers to link blocks in a chain. A hash pointer is a message digest calculated from the information content of a block. Whenever a node attempts to tamper the information content, a small change in the information leads to a ripple effect of changes in hash-pointers — making it impossible to change the information once it has been recorded in the blockchain. 

Facilitating mediator-less business transactions and supporting immutability of stored information are the two significant quality attributes associated with blockchain-based information systems. These quality attributes are derived from replicating ledgers on all the nodes in a network and linking blocks in a chain through hash pointers.

\subsection{Cryptography}\label{crypt}
Blockchain technology makes heavy use of cryptographic functions to assure trust among the users transacting over a blockchain-based business network. A typical business network includes many un-trustworthy elements.  These cryptographic functions address various purposes. Some of them are:

1) {\em Authenticating the identity of agents involved in a business transaction}:\newline Blockchain-based systems use a kind of asymmetric key cryptography. These protocols use two different keys called public and private keys. The public keys are open and used as addresses for performing business transactions while private keys are secret and used for validating the transactions. SHA-256 (e.g., Bitcoin) and ECDSA (e.g., Hyperledger) are some of the cryptographic protocols used for this purpose. 
Cryptographic functions such as digital signature are also used to authenticate a particular transaction.

2) {\em Ensuring Privacy}: The blockchain technology adopts various mechanisms to preserve the privacy of a transaction.  Below we discuss these mechanisms and their intentions behind the design.  
\begin{enumerate}
    \item  {\em Decentralised Privacy}.   The blockchain technology adopts decentralization as one of the guiding design principles. It eliminates the role of mediator to store transaction information at a central place.  The transaction information is distributed throughout a business network. Thus the threat of a mediator sharing the transaction information with a third party is eliminated.  
\item {\em Use of Asymmetric Cryptography}.  The blockchain technology uses asymmetric key cryptography to protect the identity of transaction owners and to authenticate a transaction.   Transactions are delinked from the real-world identity of transaction owners. The transaction owners are identified through using public keys which an owner can generate multiple times.  In this way, transactions are pseudo-anonymous. The private keys are used to authenticate a transaction. 
\item  {\em Additional Mechanism for Anonymity}:  In the majority of blockchains implementations, transaction owners are identified through pseudo-anonymous identity. To provide full anonymity, additional mechanisms such are mixing transaction information, and a cryptographic technique called Zero-Knowledge proof can be used.  In zero-knowledge proof, is a verification technique which assures the validity of information without disclosing additional information.
\end{enumerate}

\begin{table}[]
    \centering
    \begin{tabular}{|p{0.5in}|p{1in}| p{1.5in}|p{1.5in}|}
    \hline 
        Sr. No.  &   Consensus Algorithm & Tolerated power of adversary. & Throughput \\ \hline 
          \multicolumn{4}{|p{4.5in}|}{\center{\bf Public Blockchain}}      \\ \hline 
          
     1 & PoW & 50\% &  Low \\ \hline 
     2 & PoS & 50\% &  Good \\ \hline 
      \multicolumn{4}{|p{4.5in}|}{\center{\bf Private Blockchain}}      \\ \hline 
          
     1 & Paxos/Raft & 50\% &  Good \\ \hline 
     
      \multicolumn{4}{|p{4.5in}|}{\center{\bf Consortium Blockchain}}      \\ \hline 
          
     1 & PBFT & 33\% &  Low \\ \hline

    \end{tabular}
    \caption{Comparison of Consensus Algorithms \cite{zhang2019security}}
    \label{cp}
\end{table}

\subsection{Consensus Protocols}
In decentralized systems, agreeing upon the global state of the transaction is a challenge. In a centralized system, this is not an issue because only one copy of transaction history is present at the central authority (e.g., Banks main Server machine). Blockchain being a decentralized system, holds multiple replicas of transactions at several nodes. Agreeing upon the unique state of the transaction is an issue which is solved by executing a consensus process involving all the nodes in the system. This process is typically carried out in three stages. In the first phase, a node is elected/selected as a leader node to decide upon a unique state. In the second stage, transactions are validated. In the third stage, transactions are committed. A variety of consensus algorithms exists in blockchain-based system. These are often compared based upon how scalable the algorithm is and several malicious nodes it tolerates. The Proof-of-Work (PoW) algorithm used in Bitcoin is one example of the consensus protocol. It selects the leader node responsible for deciding upon a global state by solving a cryptographic puzzle. It takes about 10 minutes for solving the puzzle requiring extensive computational work and much electric energy. It can work in the presence of 50\% of malicious nodes in the network.

The Proof-of-Stake (PoS) is another consensus protocol in which a leader is selected with the highest stakes in the network. It has been found as scalable as compared to PoW, and it also works in the presence of 50\% of malicious nodes in the network.

The Practical Byzantine Fault Tolerant (PBFT) is the third example of consensus protocol which has been found scalable and works in the presence of 33\% (1/3) malicious nodes in the network.

Table \ref{cp} compares various consensus protocols used in private, public, and consortium blockchain. 
\vspace{-0.25in}
\subsection{Smart-Contracts}
Smart-contracts are the most significant element in the blockchain-based system because it provides configuring the behaviour of such systems. Blockchain programmers can customize the working of blockchain systems by writing programs called {\em Smart-Contract}. The smart contracts are scripts which are executed when a specific event occurs in a system. For example, in the context of Bitcoin, a coin may be released when more than one signatures are validated, or when miners solve a cryptographic puzzle.

These scripts can be written in a native language provided by blockchain systems or general-purpose programmable language. For example, Bitcoin provides a simple and less expressive native language to write a smart contract while Ethereum provides a Turing complete native language called Solidity to write smart contracts. In Hyperledger, blockchain programmers can write a smart contract in a general-purpose language such as Java/Go.
 
\section{Blockchain based Security Solutions}

This section describes blockchain-based approaches used to provide the solutions for the generic security requirements identified in Section \ref{genreq} in context of fog or cloud computing.

\subsection{Blockchain based Authentication} 
In a networked system such as cloud and fog environment, two modes of authentications exist. These are {\em centralized authentication}, and {\em decentralized authentication}. For example, OAuth 2.0 is a centralized authentication protocol. In such protocols,  a centralized authentication server verifies the credentials submitted by a client, and it authorizes to access the third party the requested services when it successfully validates the client. Majority of cloud service providers adopt this mode of authentication. Authentication services from Google, Facebook, and Twitter act as authentication servers with the login id and password on these platforms play the role of the client's credentials. Such kind of centralized authentication servers suffer from a single point of failure, and it also invades the privacy of clients \cite{almadhoun2018user}.

Decentralized authentication protocols overcome the limitations of a centralized scheme. Pretty Good Privacy (PGP) and Web of Trust (WoT) are some of the examples of decentralized protocols.  Blockchain technology is a platform supporting decentralized application development. Hence, it facilitates the development of decentralized authentication services. This section reviews some of the techniques that use blockchain technology for authentication purpose. 

Fog systems or IoT use blockchain technology to implement in many ways.  In the first kind of implementation,  Fog nodes authenticate a client or edge device through a {\em smart-contract} running on the fog nodes. The smart-contract stores a  mapping of edge devices and authorized users along with their credentials. Upon the receipt of an authentication request, the smart contract running on any of the Fog nodes can validate the submitted credentials \cite{almadhoun2018user}. 

In the second kind of blockchain-based authentication protocol, the system makes use of {\em distributed ledgers} for storing credential information and authorized device mapping.  Typically the credential information includes asymmetric key cryptography or digital signatures. Any fog node running blockchain instance known as miners can authenticate a request to access the desired service \cite{moinet2017blockchain}.

In the third kind blockchain-based variant, edge devices are grouped into a cluster called {\em bubbles of trust}. The edge devices can send/receive messages within the {\em bubbles- of- trust}.  A master node administers each bubble-of-trust.  A request for send or receive is a transaction to be recorded in the blockchain.  The master node validates a send/receive request similar to the case of a certification authority \cite{hammi2018bubbles}.

The blockchain-based authentication mechanisms have been evaluated for various kinds of security threats, and they are found robust for denial of service attacks. Also, these protocols have been scalable as compared to centralized ones\cite{hammi2018bubbles,almadhoun2018user,moinet2017blockchain}.

\subsection{Blockchain based Secured Communication}
The Fog/Cloud systems which adopt blockchain technology to implement authentication services also use the same for secured communication. As discussed in Section \ref{crypt},  the blockchain technology uses cryptography algorithms to communicate between nodes and to store data in distributed ledgers. 

As seen earlier in Section \ref{genreq}, two kinds of communication need to secure: (i) from an edge device to a fog node and,  (ii)between one fog node to another fog node. Typically blockchain is implemented as a fog service running on fog nodes.

The communication between edge devices and fog nodes (i.e. blockchain service ) is secured by assigning a public address. In the case of Ethereum,  an edge device is identified through a 20-byte address. This address can be leveraged to establish an SSL session between an edge device and a fog node \cite{almadhoun2018user}.  
By default, all the communications between fog-nodes use asymmetric-key cryptography. 

The blockchain system adopting secured communication have been found resilient to attacks such as man-in-the middle and replay attacks. Thus ensuring data confidentiality, data integrity and communication integrity. 

\subsection{Blockchain based Availability}

 Distributed ledgers and smart contracts are the two storage and computational elements in a blockchain-based system.  Multiple copies of these elements exist throughout the blockchain network.  Consensus protocols maintain a consistent global state of storage and computational elements.   Because of these inherent design properties, blockchain-based fog services are resilient to a single-point failure. Hence, they are fault-tolerant, thus reducing down-time.

Denial of service attacks is another means to disrupt the functioning of fog services. A blockchain-based system can adopt hierarchical mechanisms to defend itself from such an attack. One such mechanism is implemented in \cite{dorri2017blockchain}.   At the device level, blockchain miners protect the edge devices against deploying malware on edge devices by malicious users.  Because all miners authorise and validate an access to edge devices. At the network level, it is the responsibility of blockchain  to validate each communication emanating from edge devices and among the fog nodes. Further, as explained in \cite{hammi2018bubbles}, a blockchain-based system can dynamically form {\em bubbles-of-trust} to limit send/receive operations with a group of trusted edge devices or to isolate a malicious node/device.

\subsection{Blockchain based Privacy}
Cloud/Fog service developers who adopt {\em client-server} model for interaction have a limited set of primitives (e.g., storing personal information in the encrypted format) at their disposal to protect the privacy of the personal information shared by their users.  Unlike this,  blockchain, is a peer-to-peer system, provides a range of mechanisms to protect the privacy of personal data. Below, we explain some of these primitives.
\begin{enumerate}
    \item {\em Pseudo-anonymity} Blockchain-based system facilitates de-linking of user's real-life identity from its system identity. A user can use as many public keys as s/he wishes to perform an interaction. Also,  s/he can use a hash of some of its real-time identity for performing an interaction.   The approach, explained in \cite{al2017medibchain},  adopts this technique to protect health records of patients.
    
    \item {\em Data Ownership} In blockchain-based system, it is possible to own and control access to the personal data by the concerned user\cite{zyskind2015decentralizing}. Unlike centralized systems, data is owned and controlled by service providers.
      \item {\em Fine-grained Authorization} Data access can be authorized at multiple levels(e.g. file, record, field) by data owners. Also, one-time data access in contrast to perpetual data access is possible to grant \cite{zyskind2015decentralizing}. 
    \item{\em Encrypted Storage} Data is always stored in an encrypted format. Data owner's public and private keys are required to decrypt the data.
    \item {\em Data Transparency} Data owner is aware of what kind of data about him is collected, and it's intended use.
    \item{\em Incentives for maintaining Privacy} The  Reference \cite{li2018creditcoin} explains an application of blockchain which forwards safety-critical information (e.g., news of an accident) in a transport system without disclosing the identity of the forwarder. System rewards such {\em good} behaviour through incentives in the form of a coin which adds to their reputation.
    \item{\em Data Provenance} It refers to maintaining metadata about the creation and each access operation performed on the data. Such kind of metadata is useful for accountability and forensics purposes, which also increases data privacy. The Reference \cite{liang2017provchain} describes an application blockchain technology for data provenance. 
    
\end{enumerate}

\subsection{Blockchain based Trust Management}

Computing trust is a challenging task. Blockchain technology provides various mechanisms to handle it. This section reviews some of the blockchain-based approaches to computing trust in decentralized systems.

First computational challenges arise from its subjectivity. Trust is subjective. To handle the subjectivity, trust is either computed for an entity or for the delivered data or in a combined way. For example in case of a vehicular ad-hoc network, the trustworthiness of a vehicle needs to be defined or trustworthiness of a received message such as a notification about road accident, or trustworthiness of both message as well as who sent it. In Reference \cite{lu2018bars}, the blockchain-based anonymous reputation system is explained which computes trustworthiness of a sender and the received  message.  Historical interactions and indirect opinions of other participating nodes are used to calculate the trustworthiness of a message and a sender.

The second computational challenge arise from the fact that trust changes over the period of time.  To address this challenge a blockchain-based solution is developed in Reference \cite{moinet2017blockchain}. The approach calculates the trustworthiness of a node, in the context of wireless sensor networks. The reputation of a node is calculated based on how it responds to an event.  A reputation factor  is associated to every event. To make it relevant with respect to time, reputation factor is a  continuously decreasing function. The immutability feature of the blockchain plays a role to assign a reputation factor to nodes based on its historic interaction.

The third computational challenge is to develop a trust model which is generic in the sense that computational process is applicable to multiple domains.  This challenge is addressed in Reference \cite{iqbal2019trust} which provides a blockchain-based solution by identifying diverse attributes for calculating trust. These attributes includes: reputation, context, environment, goals, expectations, social relationships,  willingness and timeliness of evaluation. 
The approach further demonstrates the applicability of the model in the domain of Social Internet of Vehicle (SIoV). It further states that the emerging technologies such as blockchain and fog computing are appropriate for  providing scalable solution for managing trust in the dynamic environment such as (SIoV).

\section{A Performance  Analysis of  Blockchain and Fog  Computing Integration} The blockchain computing, particularly public blockchains such as Bitcoin, is known for its high energy consumption and low throughput.  In this context, the use of blockchain technology in a resource-constrained environment such as Fog and Edge Computing is questionable.  In this section, we discuss the performance analysis of implementing blockchain as a fog service.

As seen from Table \ref{secreq}, the security requirements for the majority of the Fog Computing use cases are of the type to authenticate users, to ensure the privacy of data,  to check the integrity of data, and to provide secured communication.  

For the well-known use cases of Blockchain technology such as in the financial sector, asset management and supply-chain management, the additional security requirements are to maintain consistency of data in a decentralized network, the provenance of data and seamless execution of business processes.  To realize all such security requirements, the elements of blockchain technology, such as consensus protocol and smart contract,  in addition to distributed ledger and cryptography, are necessary to implement. In such contexts, the computation demand and energy consumption are typically high.

But, a lightweight blockchain implementation that includes minimal elements of blockchain technology such as distributed ledger and cryptography can meet the majority of the security requirements of the Fog Computing use cases. 

Further, such a lightweight implementation supports different configurable deployment options.  For example, a blockchain service can be deployed with (e.g., Cloud+Fog deployment) or without Cloud (e.g., Fog only deployment model).   Such a lightweight implementation additionally shall realize the tasks of encrypting and decrypting data in the hardware with a secured wireless protocol (e.g., Zigbee) to achieve secured communication. 

Performance of one such lightweight implementation has been reported in \cite{tuli2019fogbus}. It demonstrates the use of Blockchain technology in Fog computing context for the smart-healthcare use case. It observes that the energy consumption and latency requirement is acceptable for the health care use case even when blockchain service is implemented in the Cloud+Fog integration environment.

However, the performance of Blockchain and Fog Computing integration with various tuning parameters needs to be evaluated in other application domains of IoT.

\section{Conclusion}
Identifying security requirements for an emerging computing platform is a challenging task. In this chapter, we address this challenge in the context of Fog Computing.  The emerging paradigm of Fog computing assures to deliver reduced latency time, better throughput and increased scalability to many applications designed around resource-constrained edge devices. 

Due to this assured performance, Fog Computing is increasingly preferred over  Cloud Computing platform in various safety-critical application domains.
Few examples of such application domains include Urban Surveillance and Public Safety, Smart Grid, Geospatial Data Analysis, Intelligent Transport Systems, Smart Health care, and Industry 4.0.  

All these domains have stringent security requirements. Hence a trustworthy platform is required to process information in these domains. Despite the blockchain technology's numerous drawbacks such as high energy consumption, an evolving ecosystem of developers, and legal constraints in the deployment of blockchain-based solutions;   software developers prefer to adopt the blockchain technology as a robust platform to meet many security requirements.

The chapter describes blockchain-based solutions for authentication, secured communication, availability privacy, trust management in the context of fog computing. It assumes that the blockchain as a service is available either at the layer of Cloud or Fog. Such deployments of blockchain-based solutions have been found scalable and robust to many known security attacks.

\bibliographystyle{plain}
\bibliography{edgeComputing}

\end{document}